# Characterizing dissociative motion in time-resolved x-ray scattering from gas-phase diatomic molecules

Authors: MR Ware*^[1,2], JM Glownia*[3], N Al-Sayyad[1,2], JT O'Neal[1,2], PH Bucksbaum[1,2,4]

[1]Stanford PULSE Institute, SLAC National Accelerator Laboratory, Menlo Park, CA 94025, USA
[2]Department of Physics, Stanford University, Stanford, CA 94305, USA
[3]Linac Coherent Light Source, SLAC National Accelerator Laboratory Menlo Park, CA 94025, USA
[4]Department of Applied Physics, Stanford University, Stanford, CA 94305, USA
*These authors contributed equally
^Corresponding email: mrware@stanford.edu

**Abstract:**

Time-resolved x-ray scattering (TRXS) measures internuclear separations in a molecule following laser-induced photoexcitation [1,2]. The molecular dynamics induced by the excitation laser may lie on one or several bound or dissociative electronic states [3]. TRXS from these states can be difficult to isolate because they generally overlap in the angle-resolved x-ray scattering pattern $I(x, y, \tau)$, where $\tau$ is the pump-probe delay and $x, y$ are the physical pixel positions [4]. Here we show how standard transform methods can isolate the dynamics from individual states. We form the temporal Fourier transform, $\tilde{I}(x, y, \omega) = \int_{-\infty}^{\infty} d\tau\, e^{-i\omega\tau}\, I(x, y, \tau)$. This frequency-resolved x-ray scattering (FRXS) signal segregates the bound states according to their vibrational frequencies, $\omega_i$ [5], and also displays dissociative states along straight lines $\omega = vQ$, where the slope $v$ is the rate of increase of the internuclear distance and $Q$ is the momentum transfer between the incident and scattered x-ray photon. We derive this relation and use FRXS to extract state-specific dynamics from experimental TRXS from molecular iodine following a 520 nm pump. Dynamics observed include one- and two-photon dissociation of the $^1\Pi_u$ and $^1\Sigma_g^+$ excited states, and vibrational wave packets on the B ($^3\Pi_{0u}^+$) state.

**Introduction:**

The ability to observe time-resolved motion of electrons and nuclei in molecules is one of the principal goals of femtochemistry [6]. Time-resolved x-ray scattering (TRXS) has enabled the observation of nuclear motion in molecules [7–10] and has the potential to track electronic motion [11]. In this scheme, an x-ray probe pulse scatters off an ensemble of optically photoexcited molecules, and this process is repeated for a series of delays between the optical excitation and the x-ray probe. One advantage of x-ray scattering is that the x-rays scatter from all of the electrons in the molecular system under study and, thus, embeds spatial information about the electronic charge distribution of a molecule. This has enabled the probing of parallel and perpendicular transitions in N-methyl morpholine [12]**,** the direct observation of bound and dissociative motion in molecular iodine [8,9], and the resolving of various molecular trajectories in 1,3-cyclohexadiene [7].

TRXS, however, is not a direct probe of nuclear position. Since scattering takes place in momentum or, rather, reciprocal-space, it is necessary to either fit the data to a model [7,10] or invert the data from reciprocal-space to real-space to recover the nuclear dynamics [8,9]. The inverse problem to obtain the nuclear dynamics, i.e. the pair-distribution function, is difficult because the observed momentum transfer $\vec{Q} = \vec{k}_0 - \vec{k}_s$ between the incoming and outgoing x-ray photon is limited to a few $\text{Å}^{-1}$ at an x-ray free-electron laser (FEL) [13]. For example, the scattered x-rays in a typical experiment at the LINAC Coherent Light Source (LCLS) are restricted between $Q_{min} =$ 0.1-1 $\text{Å}^{-1}$ and $Q_{max} = \frac{2\pi}{\lambda} \approx$ 4.5 $\text{Å}^{-1}$ [14]. This restricts the spatial resolution to $\Delta R = \frac{2\pi}{Q_{max} - Q_{min}} > 1.4$ Å. There is, however no comparable technical restriction on the pump-probe delay, $\tau$, which may range over a long time $\tau_R$ with fine steps $\Delta\tau$. A temporal Fourier transform of the scattering pattern $I(x, y, \tau)$ forms the frequency-resolved x-ray scattering signal (FRXS) $\tilde{I}(x, y, \omega)$, where the maximum observable beat frequency is given by $\frac{2\pi}{\Delta\tau}$ and the frequency resolution is $\frac{2\pi}{\tau_R}$. The temporal Fourier transform thus generates a high-resolution representation of TRXS.

Temporal Fourier transform methods have been developed for Fourier Transform Inelastic X-ray Scattering (FTIXS), to characterize phonon modes in crystals to arbitrary frequency resolution [15]. Temporal transform methods for TRXS from gases have been used to isolate diatomic vibrations [5].

Here the expression for FRXS of a dissociating diatomic molecule is derived, and the frequency spectrum is shown to isolate dissociative motion along lines in reciprocal-space and reciprocal-time, $Q$ and $\omega$, despite dissociation not being periodic in real-space, i.e. dissociations follow a trajectory like $R(\tau) = R_0 + v\tau$. The analytical results are then confirmed through comparison to measured experimental data, where two dissociative states are observed: one parallel to the pump laser and one perpendicular.

**Experiment:**

The experimental apparatus to study the time-resolved x-ray scattering of molecular iodine following photoexcitation by 520 nm light has been described elsewhere [5,8,14] and a short description and schematic is provided in Figure 1. An attempted real-space reconstruction of the data was published previously [8]. In contrast, the analysis in this paper using FRXS highlights a previously undescribed approach to isolate and characterize dissociations measured using TRXS.

The specifics of the experiment are as follows. A 50 fs, 20 uJ, vertically polarized pulse of 520 nm light is focused into a gas cell containing around 50 Torr of molecular iodine at ~100$^{o}$C. Following the optical pump, a 50 fs, 2 mJ, horizontally polarized x-ray probe arrives at the scattering cell with variable delay. The resulting scattering is measured in the forward direction by the Cornell-SLAC Pixel Array Detector (CSPAD) [13]. Scattering is measured at each pump-probe delay, resulting in the difference image shown in Figure 2.

At 520 nm, one photon will excite molecular iodine to the bound B ($^3\Pi_{0u}^+$) state or the dissociative $^1\Pi_u$ state. Two photons can access higher dissociative states, including the $^1\Sigma_g^+$ state. The excitation begins near the equilibrium position of the ground X ($^1\Sigma_g^+$) state at $R_0 = 2.666$ Å. The corresponding potential energy curves are shown in Figure 3 from Reference [16]. In the bound B state, $520 \pm 5$ nm light excites highly anharmonic vibrations with periods between $T = 520$ to 650 fs and angular frequencies between $\omega = 9.6$ to 11.9 THz. In the dissociative $^1\Pi_u$ state, the internuclear separation increases at a rate of $v = 16$ Å/ps, and in the dissociative $^1\Sigma_g^+$ state, $v = 20$ Å/ps. The beat frequencies and periods were derived from [17] assuming a pump wavelength of 520 nm, and velocities can be derived from the potential energy curves from [16]. The initial position, $R_0$, the above frequency, $\omega$, and velocities, $v$, are observed using FRXS, which will now be described.

**Theory:**

As discussed in previous papers [2,5,18], time-resolved x-ray scattering may be expressed as a product of three factors

$$\frac{dI}{d\Omega} = \frac{d\sigma_{Th}}{d\Omega} I_0 \langle F(\vec{Q},\tau)\rangle, \quad (1)$$

where $\frac{dI}{d\Omega}$ is the number of photons scattered into a solid angle $\Omega$, $\frac{d\sigma_{Th}}{d\Omega}$ is the Thomson scattering cross section, and $I_0$ is the incident x-ray intensity. $\langle F(\vec{Q},\tau)\rangle$ is a time- and angle-dependent polarization-corrected scattering probability given by

$$\langle F(\vec{Q},\tau)\rangle = 2|f_A(Q)|^2\left(1 + S(\vec{Q},\tau)\right) \qquad (2)$$

for a homonuclear diatomic molecule. In Equation 2, $f_A(Q)$ is the atomic form factor and $S(\vec{Q},\tau)$ is the molecular scattering factor, which encodes the internuclear separations. This expression is correct in the limit where the independent atom approximation holds, i.e. heavy atoms and insufficient time-resolution to observe coherent effects between electronic states [18,19]. Within this approximation, $S(\vec{Q},\tau)$ may be considered for each electronic state independently and may be expressed as

$$S(\vec{Q},\tau) = \int d\vec{R}\rho(\vec{R},\tau)\cos\left(\vec{Q}\cdot\vec{R}\right), \qquad (3)$$

where $\rho(\vec{R},\tau)$ is the internuclear probability density on some electronic state [2,20].

Turning to FRXS, consider a perfectly aligned classical diatomic molecule, which dissociates along a trajectory $\vec{R}(\tau) = (R_0 + v\tau)\,\hat{e}_z$ for $\tau > 0$. For this special case, the molecular scattering factor from Equation 3 evaluates to

$$S(Q_z,\tau) = \cos Q_z R_z(\tau). \qquad (4)$$

Now, FRXS is defined as

$$\tilde{S}(\vec{Q},\omega) = \int_{-\infty}^{\infty} d\tau\, e^{-i\omega\tau} S(\vec{Q},\tau). \qquad (5)$$

Therefore, for the above special case in Equation 4, the FRXS signal is

$$\tilde{S}(Q_z,\omega) = \frac{1}{2}\left(e^{\frac{i\omega R_0}{v}}\delta(\omega - vQ_z) + e^{\frac{-i\omega R_0}{v}}\delta(\omega + vQ_z)\right). \qquad (6)$$

The above equation has two important properties: (1) the maxima of the FRXS lies along $\omega = vQ$ and (2) the phase evolves like $\phi = \omega R_0/v$. The rate of increase of internuclear separation in a dissociation may then be obtained by fitting the position of the maxima to a line, and the initial position of the dissociation may be obtained by fitting the phase along those maxima to another line, as will be demonstrated in the discussion section.

Before turning to the experimental analysis, the physical alignment of a diatomic will now be considered. Following excitation by a polarized laser pulse, the angular distributions will go as $\cos^{2n}\theta$ for parallel transitions or $\sin^{2n}\theta$ for perpendicular transitions, where $n$ indicates the number of photons absorbed and $\theta$ is the angle with respect to the laser polarization axis as shown in Figure 1. These distributions may be expressed as a linear combination of Legendre polynomials, $P_l(\cos\theta)$, such that the molecular scattering factor from Equation 3 may be rewritten as

$$S(\vec{Q},\tau) = \sum_l P_l(\cos\theta) S_l(Q,\tau), \quad (7)$$

where

$$S_l(Q,\tau) = \int dR\, R^2 \rho_l(R,\tau) j_l(QR), \qquad (8)$$

$\rho_l(R,\tau)$ is the projection of the nuclear probability function onto a given Legendre polynomial, and $j_l(QR)$ are the spherical Bessel functions. Now the FRXS may be considered for a given Legendre order $\tilde{S}_l(Q,\omega) = \int_{-\infty}^{\infty} d\tau\, e^{-i\omega\tau} S_l(Q,\tau)$. As shown in Appendix A, the FRXS for a dissociating diatomic is approximately given by

$$\tilde{S}_l(Q,\omega) \approx \frac{e^{i\omega R_0/v}}{2iQv}\left[E_1(-i(QR_0 - \omega R_0/v)) - E_1(i(QR_0 + \omega R_0/v))\right] \qquad (9)$$

for each Legendre order, where $E_1(z) = \int_z^{\infty} dt\, e^{-t}/t$ is an exponential integral. To show that Equation 9 shares the same important features as the first derivation in Equation 6, the result is shown in Figures 4b and 4c, which demonstrate that (1) the maxima of the FRXS lies along $\omega = vQ$ and (2) the phase evolves like $\phi = \omega R_0/v$.

**Discussion:**

To validate the theory derived above, a comparison to the experimental data is now made. For each pump-probe delay of the TRXS as shown in Figure 2, the Thomson cross-section and atomic form factors are divided out, $I(x,y,\tau) \to S(x,y,\tau)$, and the pixel coordinates $(x,y)$ are mapped onto the scattering coordinates $(Q,\theta)$ using the method described in Appendix B, $S(x,y,\tau) \to S(Q,\theta,\tau)$. Following the coordinate mapping, $S(Q,\theta,\tau)$ is projected onto the zeroth through tenth Legendre polynomials to obtain $S_l(Q,\tau)$. $S_0(Q,\tau)$ and $S_2(Q,\tau)$ are shown in Figures 5a and 6a respectively.

These Legendre projections are then used to generate the FRXS, $\tilde{S}_0(Q,\omega)$ and $\tilde{S}_2(Q,\omega)$, through a discrete Fourier transform (DFT). The power spectrum following the DFT is shown in Figures 5b and 6b. The power spectrum allows for the identification of the bound state and two dissociations. The bound state is peaked at $\omega = 11.6 \pm 1.1$ THz, the first dissociation has a final velocity of $16.4 \pm 0.2$ Å/ps, and the second dissociation has a final velocity of $19.9 \pm 0.2$ Å/ps. These results align with the inferred values for the B, $^1\Pi_u$, and $^1\Sigma_g^+$ states respectively, as derived from Reference [16]. For information on why the bound state appears at its beat frequency in the power spectrum, please see [5] for details on characterizing bound state motion using FRXS. The supplementary material provides a reconstruction of the B state equilibrium separation, which is found to be $3.7 \pm 0.1$ Å as compared to $3.6 - 3.8$ Å, depending on the pair of vibrational states in the anharmonic potential generating the scattering.

The dissociations can also be observed in the original images on the detector following the temporal Fourier transform, $\tilde{I}(x,y,\omega)$. The dissociations present themselves as outward moving crescents on the detector image in reciprocal-space and reciprocal-time. The slower dissociation appears first on the detector perpendicular to the pump polarization as $\omega$ is increased, where the perpendicular alignment has been seen elsewhere [21,22]. The faster dissociation then appears moving parallel to the pump polarization. This effect is shown in Figures 7a and 7b as well as a GIF included in the supplementary materials.

The faster dissociation as seen in the Legendre 2 projection in Figure 6 will be used to demonstrate that FRXS can characterize the dissociation velocity and initial position. To obtain the dissociation velocity, the position of the maximum at each momentum transfer, $Q$, is extracted from the power spectrum in Figure 6b. The positions of the maxima, $(Q,\omega)$, are then used to fit the line $\omega = vQ$ as shown in Figure 8a. This method obtains a velocity of $v = 19.9 \pm 0.2$ Å/ps as compared to the predicted 20 Å/ps from the dissociative $^1\Sigma_g^+$ state. Now using the measured velocity, the phase along the line $\omega = vQ$ is extracted to find the initial position, $R_0$. The phase as a function of the angular frequency is fit to $\phi = \frac{\omega R_0}{v} + \phi_0$ to obtain the initial position as shown in Figure 8b. (For reference the real part of the frequency spectrum is shown in Figure 6c). This method obtains an initial position of $R_0 = 2.3 \pm 0.4$ Å as compared to the known value of $2.666$ Å.

**Conclusion:**

An analysis method for leveraging FRXS to characterize dissociative motion has been derived and applied to experimental data. The obtained values for initial position and dissociation velocities align with the expectations for the $^1\Pi_u$ and $^1\Sigma_g^+$ states of molecular iodine. The novel aspect of this approach is that an interpretable and compact representation of the experimental measurement may be obtained in reciprocal-space and reciprocal-time without the difficulty of inverting the measurement to the traditional space and time representation. Thus, FRXS presents an alternative to traditional analyses of TRXS. The traditional approach is limited by the range of momentum transfer, $Q$, that is accessible at FELs. FRXS does not suffer this limitation, and in fact, FRXS leverages the strengths of FELs, namely fine time resolution and fast data accumulation. This enables a long range of pump-probe delays to be measured in an experiment, thereby improving the frequency resolution of an experiment, while maintaining sufficient temporal resolution to measure high beat frequencies. These advantages have been leveraged to obtain compact representations of dissociations along lines in reciprocal-space and reciprocal-time, demonstrating an alternative to traditional analyses of time-resolved x-ray scattering for gas-phase photochemistry.

**Acknowledgements:**

This research is supported through the Stanford PULSE Institute, SLAC National Accelerator Laboratory by the U.S. Department of Energy, Office of Basic Energy Sciences, Atomic, Molecular, and Optical Science Program. Use of the LINAC Coherent Light Source (LCLS), SLAC National Accelerator Laboratory is supported by the U.S. Department of Energy, Office of Basic Energy Sciences under Contract No. DE-AC02-76SF00515.

**Appendix A: FRXS of an aligned distribution**

To reproduce the exact structure of the FRXS given by a dissociation, the alignment of the molecule needs to be considered.

For example, an isotropic distribution will project onto Legendre 0 and have a molecular scattering distribution given by $S_0(Q,\tau) = \int dR\, R^2\, \rho(R,\tau) j_0(QR)$, where $j_0(QR)$ is the zeroth order spherical Bessel function. A $\cos^2\theta$ distribution will project onto both Legendre 0 and 2 and have a molecular scattering distribution with both the $S_0(Q,\tau)$ component as well as $S_2(Q,\tau) = \int dR\, R^2\, \rho(R,\tau) j_2(QR)$, where $j_2(QR)$ is the second order spherical Bessel function.

For simplicity, consider an aligned distribution of molecules with $R^2\, \rho(R,\tau) = \delta(R - R(\tau))$, then $S_l(Q,\tau) = j_l(QR(\tau))$ for $\tau > 0$. The frequency resolved scattering is accordingly

$$\tilde{S}_l(Q,\omega) = \int_0^\infty d\tau\, e^{-i\omega\tau}\, j_l(QR(\tau))$$

Each even spherical Bessel function contains a $\sin x\,/x$ term, which by observation makes the largest contribution to this integral. Focusing the derivation to this term, the approximate solution can be found by taking

$$\tilde{S}_{2l}(Q,\omega) \approx \int_0^\infty d\tau\, e^{-i\omega\tau} \frac{\sin QR(\tau)}{QR(\tau)}$$

Dissociation is the focus here, so $R(\tau) = R_0 + v\tau$. Then taking $u = QR(\tau)$ and expanding sine as a difference of exponentials, the integral becomes

$$\tilde{S}_{2l}(Q,\omega) = \frac{e^{i\omega R_0/v}}{2iQv}\left[\int_{QR_0}^\infty du\, \frac{e^{iu(1-\omega/Qv)}}{u} - \int_{QR_0}^\infty du\, \frac{e^{-iu(1+\omega/Qv)}}{u}\right]$$

This may be identified as the difference of exponential integrals after taking two additional u-substitutions with $a = u(1-\omega/Qv)$ and $b = u(1+\omega/Qv)$ such that

$$\tilde{S}_{2l}(Q,\omega) = \frac{e^{i\omega R_0/v}}{2iQv}\left[\int_{QR_0-\omega R_0/v}^\infty da\, \frac{e^{ia}}{a} - \int_{QR_0+\omega R_0/v}^\infty db\, \frac{e^{-ib}}{b}\right]$$

$$\tilde{S}_{2l}(Q,\omega) = \frac{e^{i\omega R_0/v}}{2iQv}\left[E_1(-i(QR_0 - \omega R_0/v)) - E_1(i(QR_0 + \omega R_0/v))\right],$$

where $E_1(z) = \int_z^\infty dt\, e^{-t}/t$ is an exponential integral.

The above equation shares the same features described in the simplified derivation in the text for perfectly aligned diatomics: (1) the maxima of the FRXS lies along $\omega = Qv$ and (2) the phase evolves like $\phi = \omega R_0/v$.

**Appendix B: Coordinate mapping**

For a diatomic excited along a polarized laser pulse, the relevant angular decomposition $(\theta,\phi)$ is shown in Figure 1. To map from the detector image $I(x,y)$ set a distance $L$ away from the scattering center onto the molecular frame $I(Q,\theta)$, as sketched in Figure 9, is a simple geometry problem. First, the momenta transfer must be decomposed in the directions parallel to the field, $\hat{e}_y$, and perpendicular to the field, the $\hat{e}_x - \hat{e}_z$ plane. The parallel decomposition is given by $Q_y = \frac{k_0 y}{\sqrt{R^2+L^2}}$, where $R = \sqrt{x^2+y^2}$ is the distance from the center of the detector, and the perpendicular decomposition is given by $Q_\perp = k_0\sqrt{\frac{x^2+L^2}{R^2+L^2} + 1 - \frac{2L}{\sqrt{R^2+L^2}}}$. With that decomposition in hand, $(Q,\theta)$ is determined by $Q^2 = Q_y^2 + Q_\perp^2$ and $\tan\theta = \frac{Q_\perp}{Q_y}$.

**Appendix C: Error propagation and data analysis**

For each shot the pump-probe delay, $\tau_i'$, is measured and then binned into some time bin, $\tau_j \pm \Delta\tau$. This allows for the generation of the mean scattered intensity at each time delay

$$I(x, y, \tau_j) = \frac{1}{N_j} \sum_i I(x, y, \tau_i'),$$

where $(x, y)$ indicate each pixel on the CSPAD detector. From these images, the unpumped signal is subtracted to find the difference scattering, $\Delta I(x, y, \tau_j) = I(x, y, \tau_j) - I_u(x, y)$.

The difference images are then divided by the Thompson cross-section $\frac{d\sigma}{d\Omega}$ [23], iodine's atomic form factor $|f_I(Q)|^2$ (23), and the correction factor for attenuation in the scattering cell [14]. This results in the difference molecular scattering factor, $\Delta S(x, y, \tau_j)$, up to an overall factor of the x-ray intensity. Now the variation at each pixel position is generated by

$$\sigma^2(x, y, \tau_j) = Var\left(\Delta S(x, y, \tau_j)\right) = \frac{1}{N_j} \sum_i \left|\Delta S(x, y, \tau_i') - \Delta S(x, y, \tau_j)\right|^2.$$

The variation is then propagated through the analysis as follows.

For the projection of $\Delta S(x, y, \tau_i')$ onto Legendre polynomials, the coordinates are first mapped from $(x, y)$ onto $(Q, \theta)$ as described in Appendix B. Then, mapping $\Delta S(Q, \theta, \tau)$ onto the Legendre coefficients, $\Delta S_l(Q, \tau)$, is achieved through a $\chi^2$-minimization. The $\chi^2$ model is defined by

$$\chi^2 = \sum_l \frac{(p_i - s_i)^2}{\sigma_i^2},$$

where $p_i$ is the fitted function, $s_i$ is the data, and $\sigma_i^2$ is the variance. For a linear model, the fitted function may be expressed as $p_i = \sum_j x_j f_j(Q_i)$, where $x_j$ are the model coefficients. Then the solution to the $\chi^2$-minimization is

$$x_k = \sum_i A_{ki} s_i / \sigma_i^2,$$

where $A = (f^T \sigma^{-2} f)^{-1} f$. The associated error for the solution, $x_k$, is then

$$\sigma_k = \sqrt{\sum_j \left(A_{kj} \sigma_j^{-1}\right)^2},$$

as shown in [24]. For the Legendre projection of $\Delta S(Q, \theta, \tau)$ onto $\Delta S_l(Q, \tau)$, the fitted function is $f_j(\theta_i) = P_j(\cos\theta_i)$, where $j = 0, 2, \dots, 10$ are used.

For the temporal Fourier transform of the Legendre coefficients, $\Delta S_l(Q, \tau)$, a discrete Fourier transform is used. As the Fourier transform is a unitary transform, the standard deviation for each frequency element is simply the sum in quadrature of the errors, $\sigma_l(Q, \tau_j)$, where

$$\sigma_l(Q, \omega_j) = \sqrt{\sum_j \sigma_l{}^2(Q, \tau_j)}.$$

**Supplemental:**

GIF of frequency-resolved scattering on the time binned CSPAD images.

**Figures:**

Figure 1:

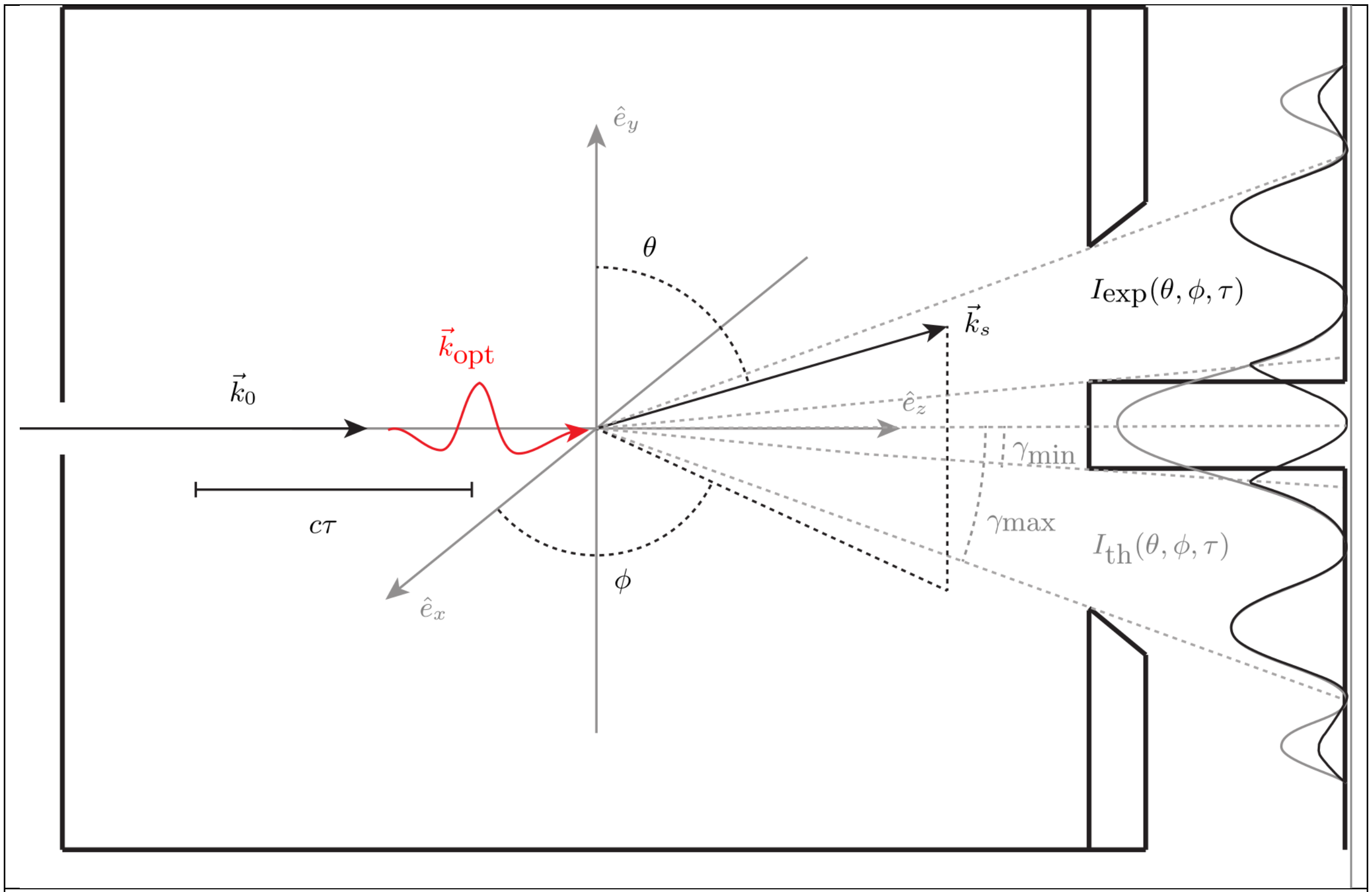

Sketch of the scattering cell and the scattering geometry of the experiment. The x-ray, $\vec{k}_0$, and pump, $\vec{k}_{opt}$, pulses copropagate into the scattering cell with a relative delay of $c\tau$. The x-rays scatter from the excited iodine molecules at some angle $(\theta, \phi)$ onto the CSPAD detector hitting a pixel $(x, y)$. Due to physical beam blocks in the gas cell, the measured signal, $I_{exp}(\theta, \phi, \tau)$, must be corrected for the attenuation at the extreme angles $\gamma_{min}$ and $\gamma_{max}$ as described in [14] to reach agreement with the theoretical result, $I_{th}(\theta, \phi, \tau)$.

Figure 2:

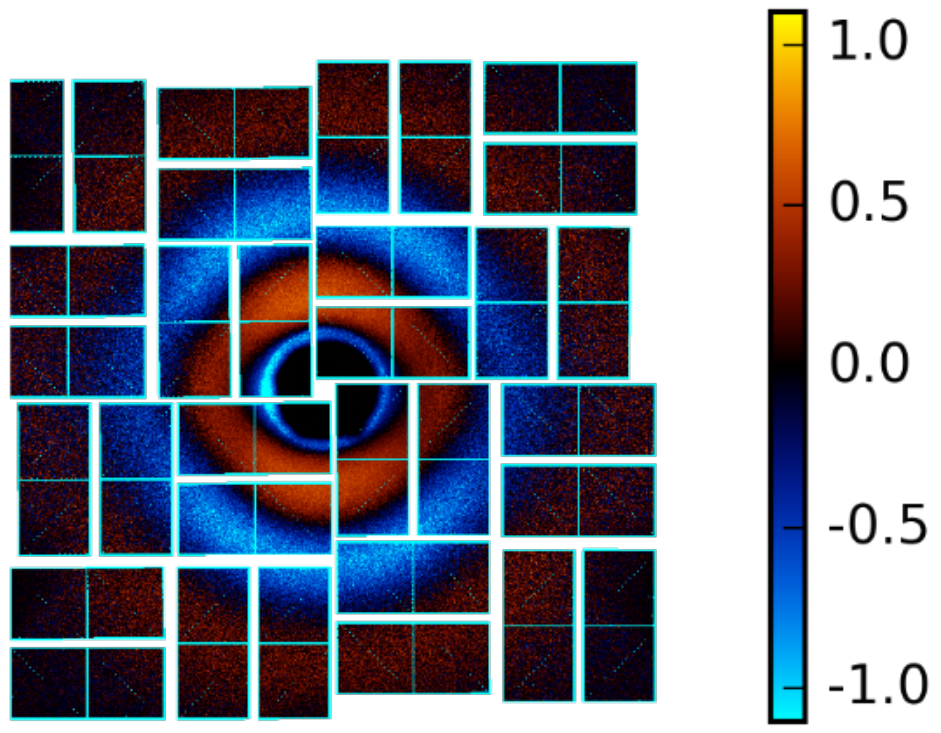

Difference image on the x-ray detector (CSPAD) for a delay of 120 fs after pumping with 520 nm light.

Figure 3:

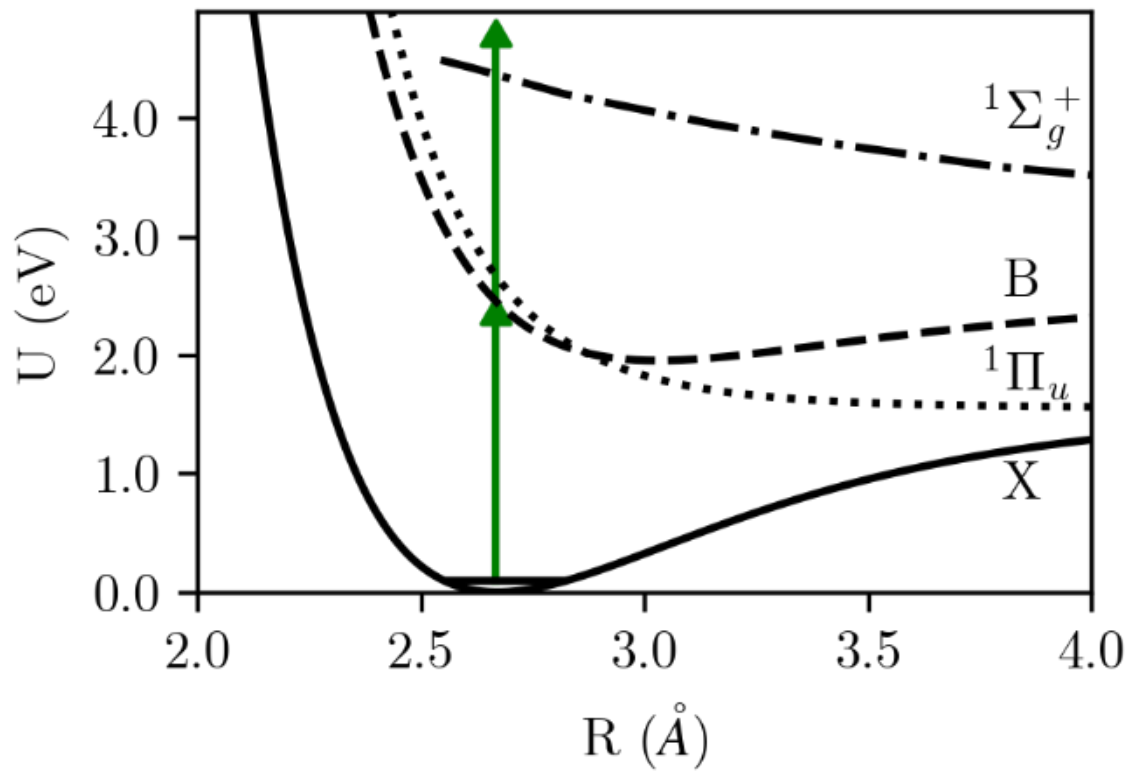

Following photoexcitation by 520 nm light, a single photon may excite high in the bound B state ($\omega =$ 9.6 to 11.9 THz) or the dissociative $^1\Pi_u$ state ($v = 16$ Å/ps). Two photons excite the dissociative $^1\Sigma_g^+$, state which shares the same symmetry as the ground X state ($v = 20$ Å/ps). These states are identified in the frequency-resolved scattering in Figures 5b and 6b. Potential energy curves from Reference [16].

Figure 4:

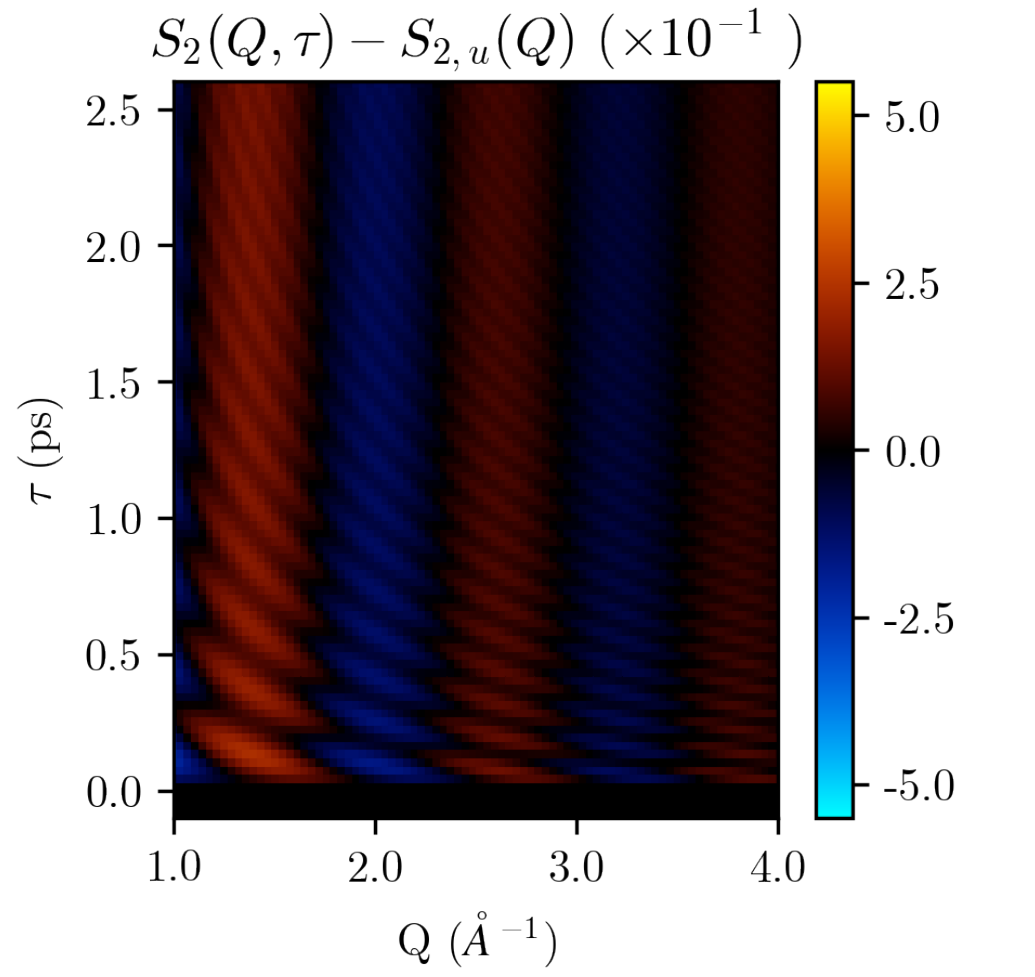

(a)

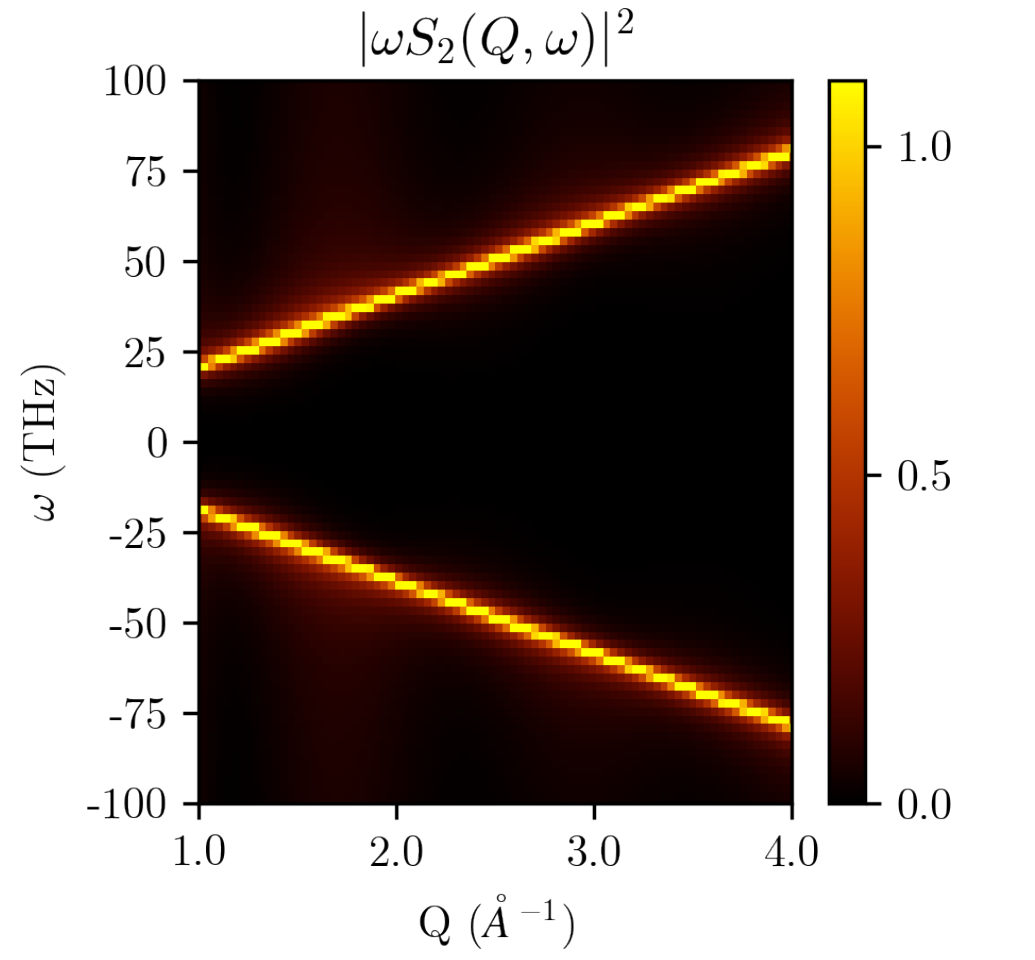

(b)

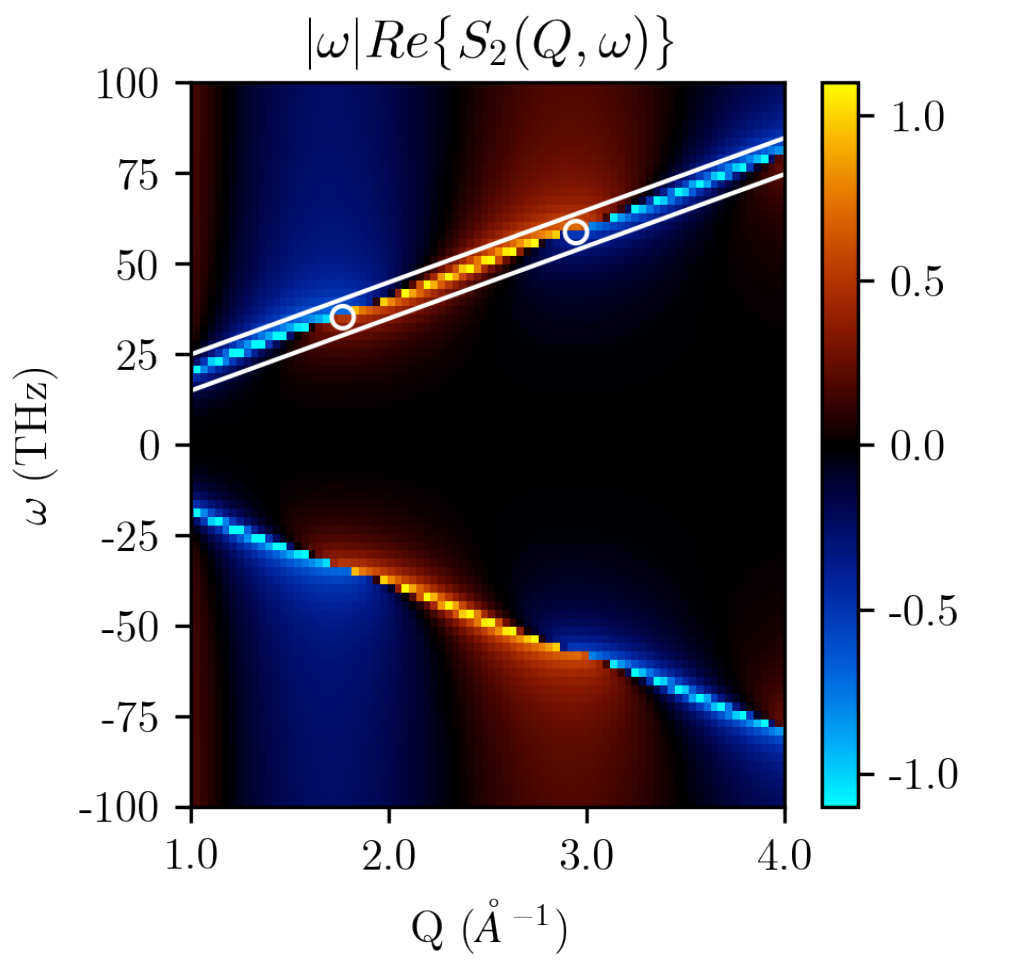

(c)

(a) Theoretical time-resolved x-ray scattering following photoabsorption of two 520 nm photons onto the dissociative ${}^1\Sigma_g^+$ state, using $v = 20$ Å/ps and $R_0 = 2.666$ Å. (b) Power spectrum under similar conditions generated using Equation 9. (c) Real part of the frequency spectrum of (b). White lines run parallel to $\omega = vQ$, and the circles identify the nodes at $\frac{\omega R_0}{v} = \frac{n\pi}{2}$, where $n = 3$ and 5. Note that a phase shift was applied to Equation 8 such that Figure 4c reproduces Figure 6c.

Figure 5:

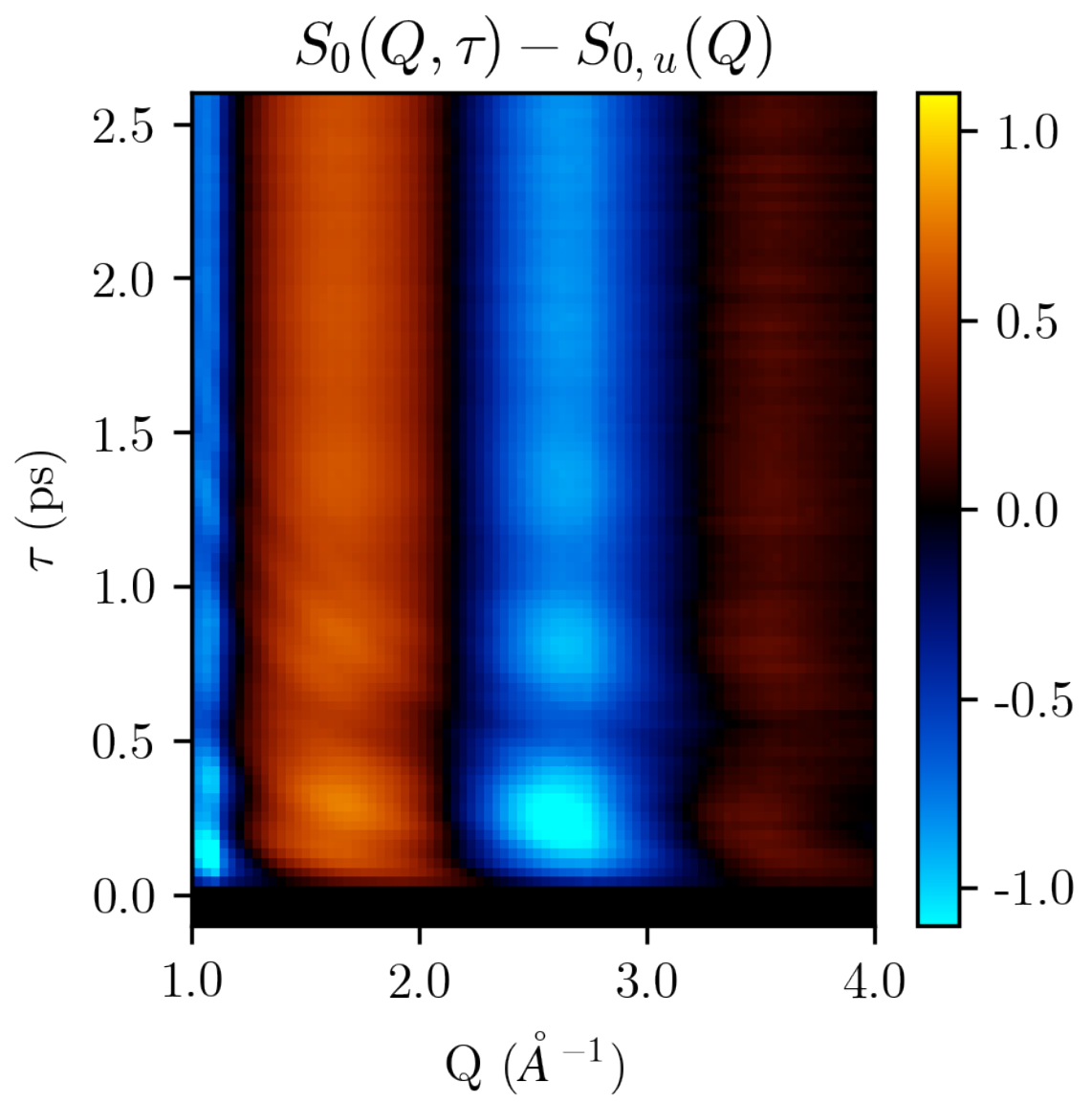

(a)

(a) Experimental time-resolved x-ray scattering following photoabsorption of 520 nm light. This is the isotropic component of the data found by projecting the CSPAD image at each time delay onto the zeroth order Legendre polynomial, $P_0(\cos\theta)$. (b) Power spectrum of (a). There are two dissociations evident in the data: $v = 19.9 \pm 0.2$ Å/ps and $v = 16.4 \pm 0.2$ Å/ps. There is also the bound B state motion peaked at $\omega = 11.6 \pm 1.1$ THz.

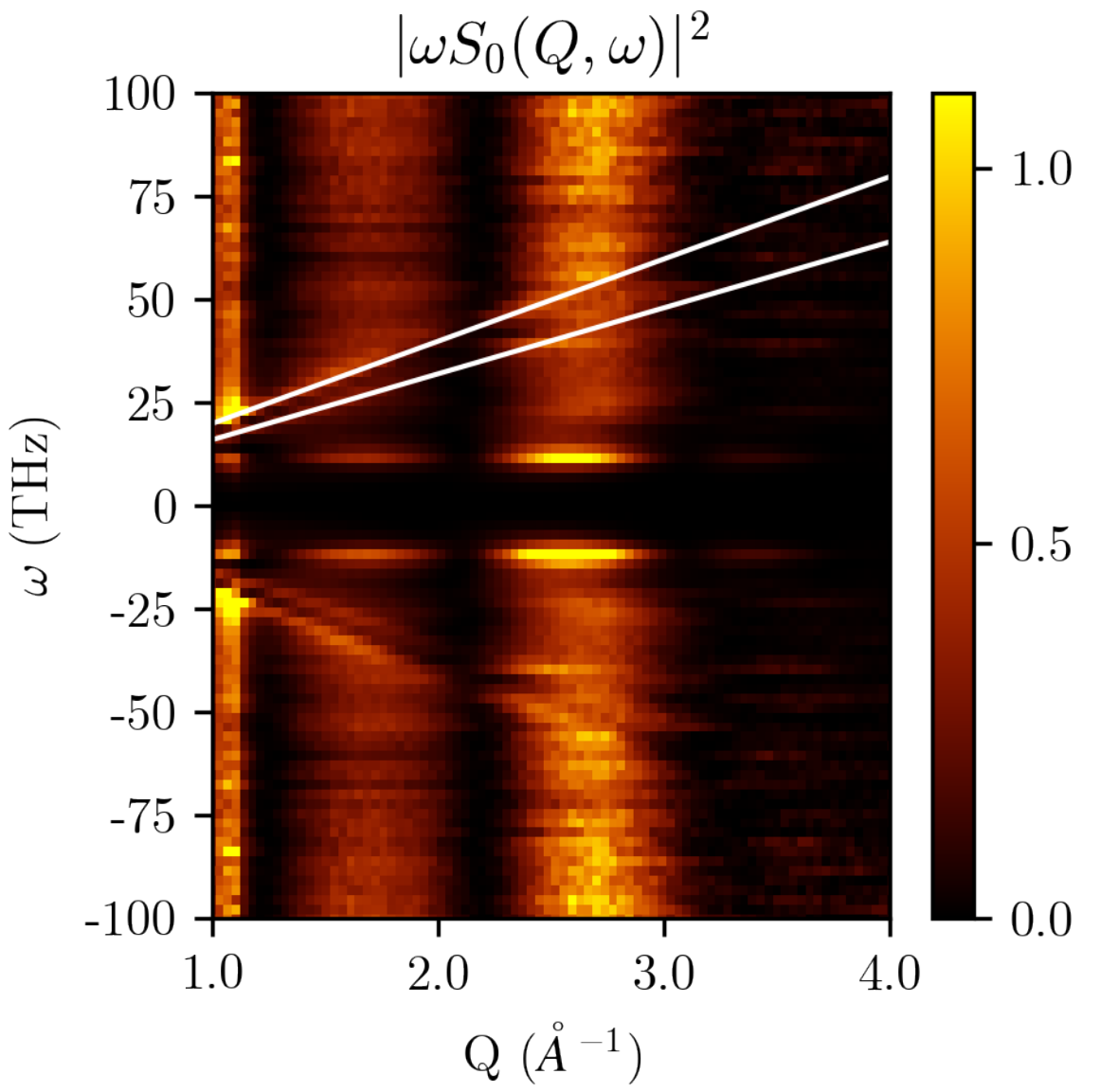

(b)

Figure 6:

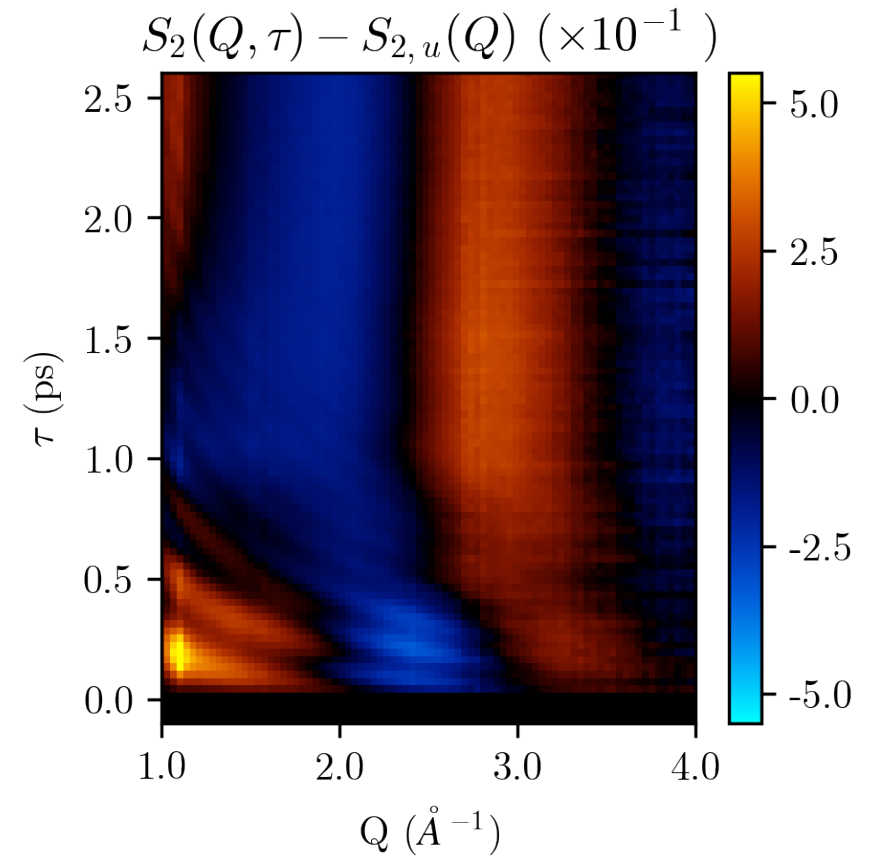

(a)

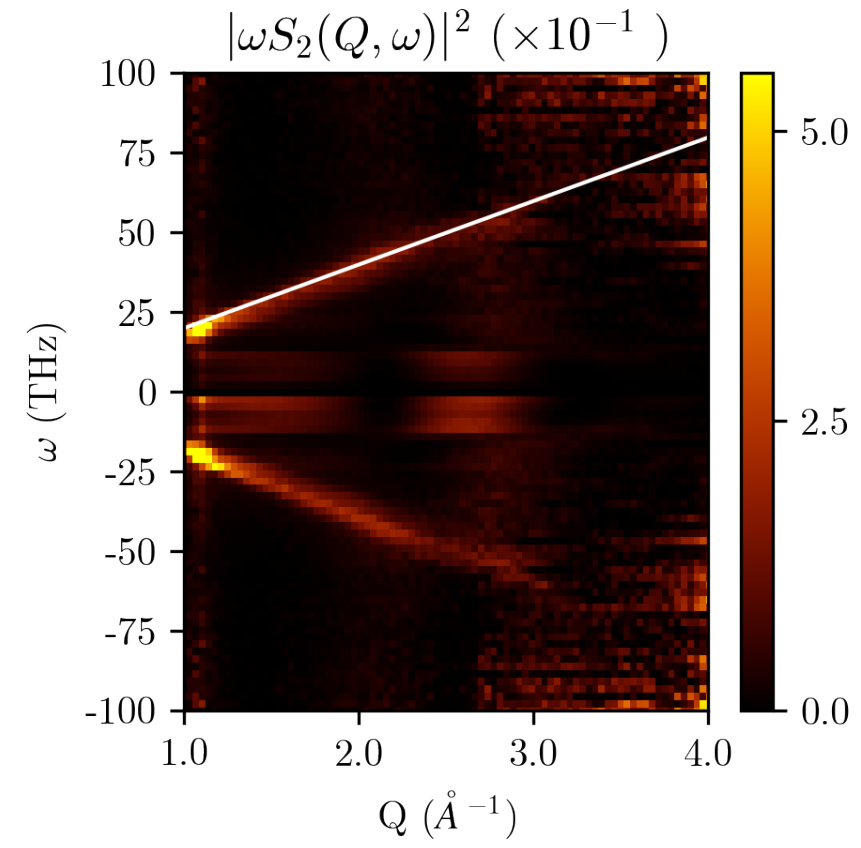

(b)

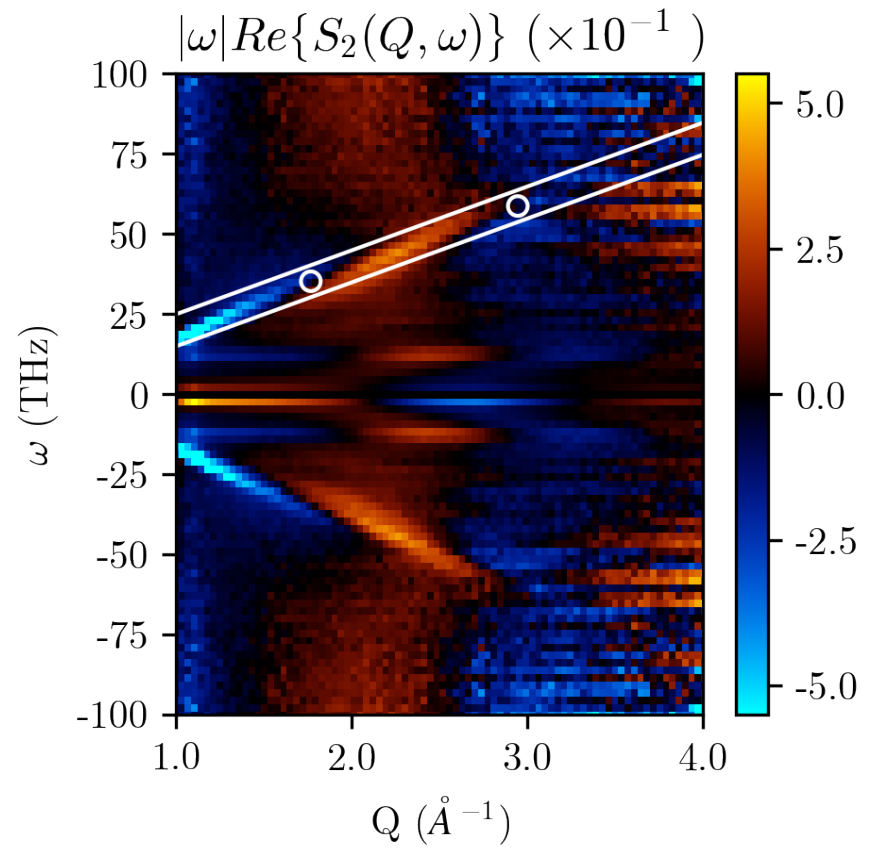

(c)

(a) Experimental time-resolved x-ray scattering following photoabsorption of 520 nm light. This is the component of the data projected onto the second order Legendre polynomial, $P_2(\cos\theta)$. (b) Power spectrum of (a). The $v = 19.9 \pm 0.2$ Å/ps dissociation is observed, and the bound B state motion peaked at $\omega = 11.6 \pm 1.1$ THz. (c) Real component of the frequency spectrum. White lines run parallel to $\omega = vQ$, and the circles identify the nodes at $\frac{\omega R_0}{v} = \frac{n\pi}{2}$, where $n =$ 3 and 5. Compare to the theoretical prediction in Figure 4c.

Figure 7:

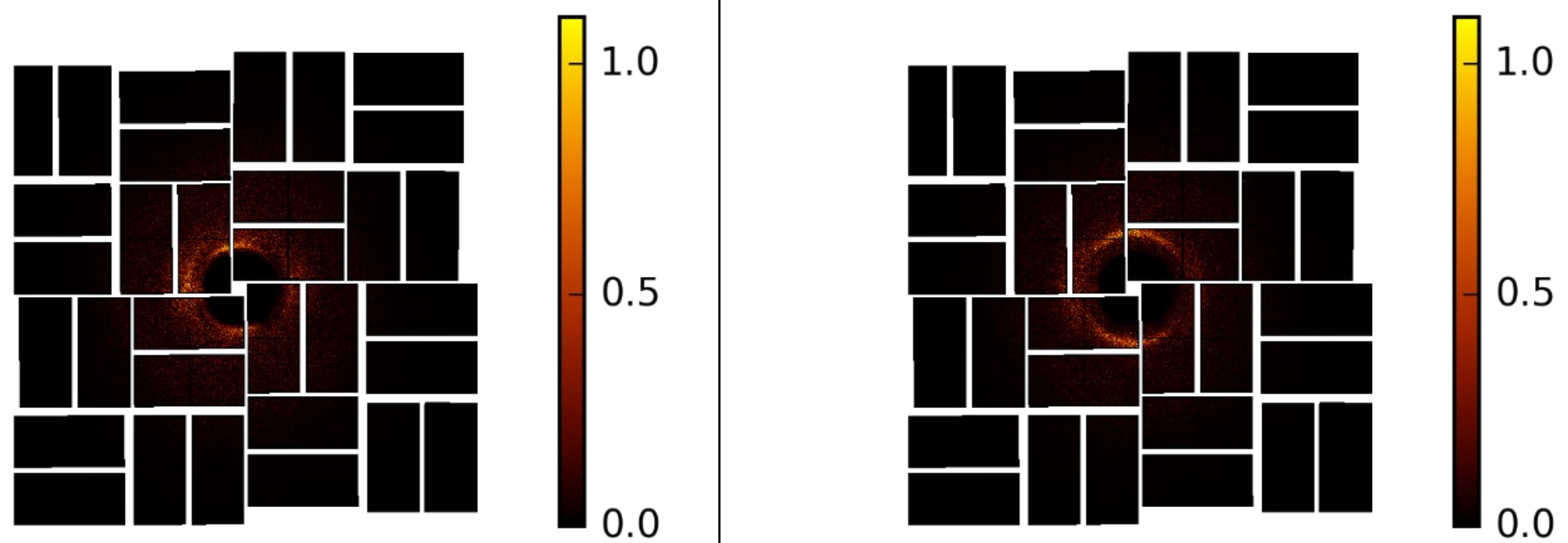

(a) (b)

Power spectrum done on the time binned detector images, $I(x, y, \tau)$, to find $\tilde{I}(x, y, \omega)$. (a) The power spectrum is shown at $\omega = 20.9$ THz across all pixels. Parallel to the laser field (y-axis) we observe the two-photon $20$ Å/ps dissociation beginning to emerge from the beam block. Perpendicular to the laser field (x-axis) we observe the one-photon $16$ Å/ps dissociation. (b, $\omega = 27.9$ THz) The two-photon dissociation has moved outward on the detector as the frequency was increased, and the one-photon dissociation is no longer visible. In the supplemental information, a GIF is included showing the two dissociations propagating outward as frequency is increased.

Figure 8:

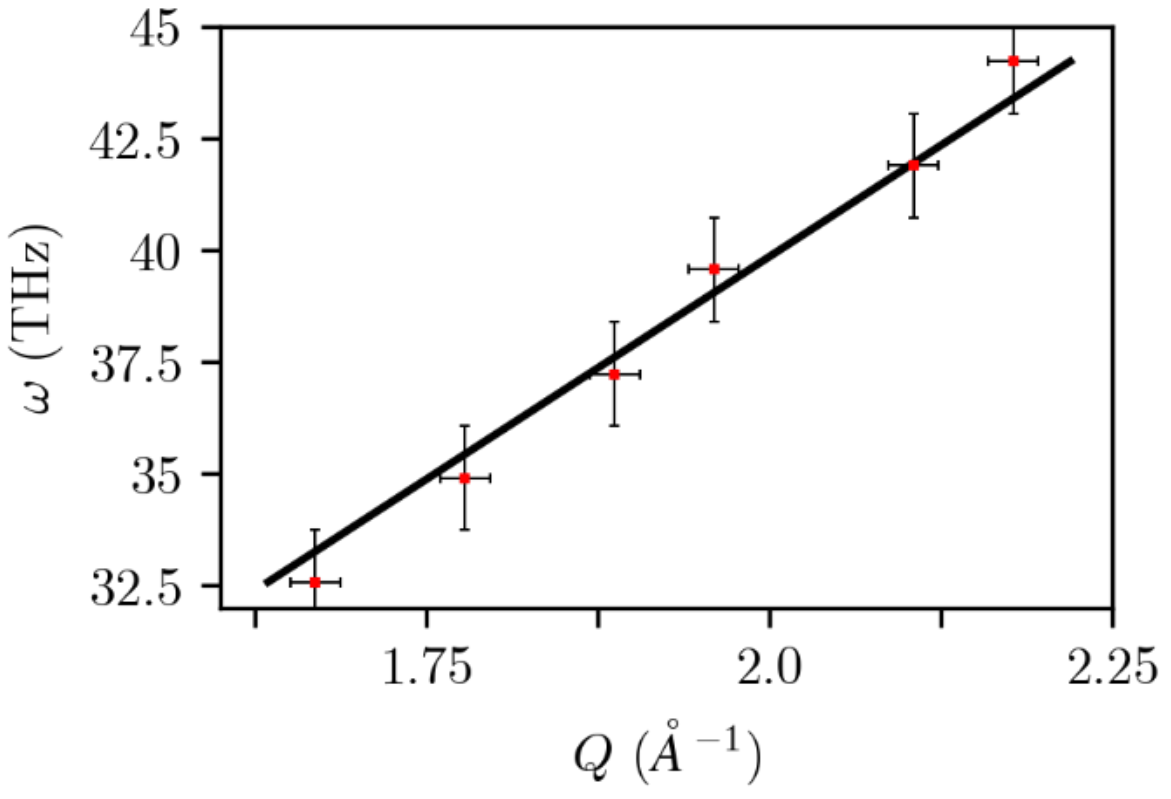

(a)

Position of the maxima at each momentum transfer, $Q$, along the dissociation line in Figure 6b shown in red. The fit to $\omega = vQ$ is shown in black, which finds $v = 19.9 \pm 0.2$ Å/ps, enabling the identification of dissociation along the $^1\Sigma_g^+$ state. Error bars are the bin size in $Q$ from Figure 6b, and error bars in $\omega$ are determined by the scan range.

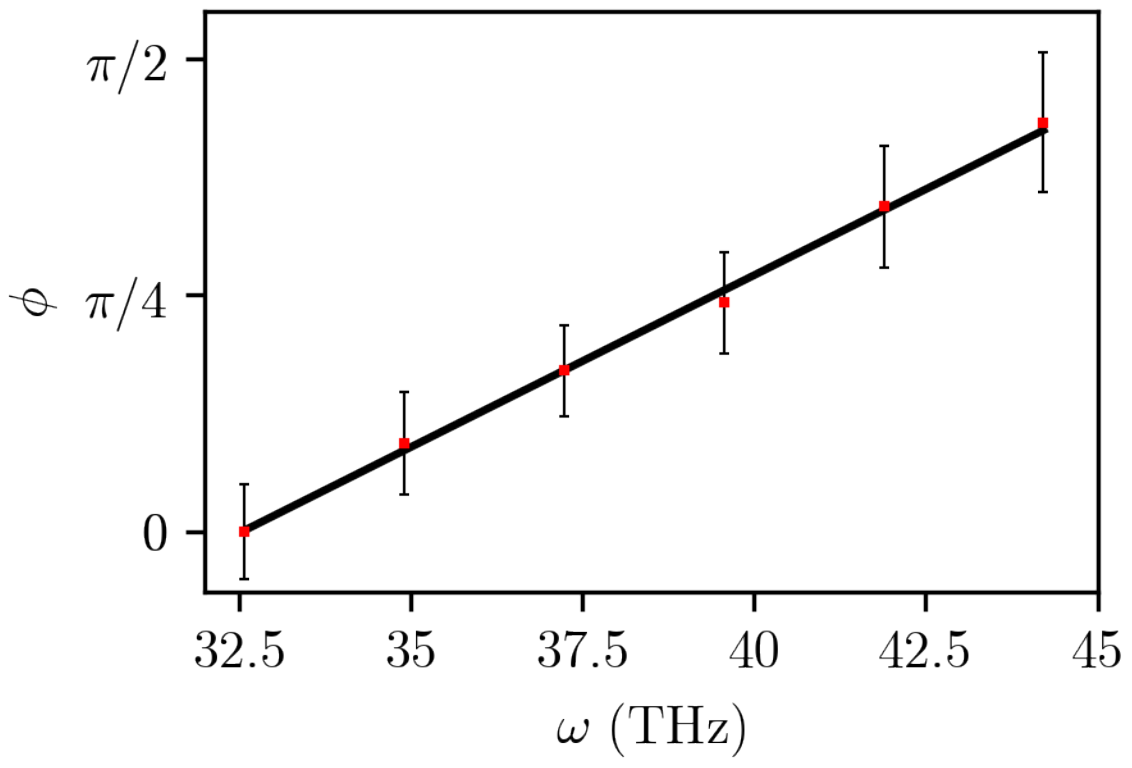

(b)

Phase along the dissociation line in Figure 6c shown in red. The fit to $\phi = \frac{\omega R_0}{v} + \phi_0$ is shown in black, which recovers the initial position before dissociation as $R_0 = 2.3 \pm 0.4$ Å. Error bars are the standard deviation of the phase.

Figure 9:

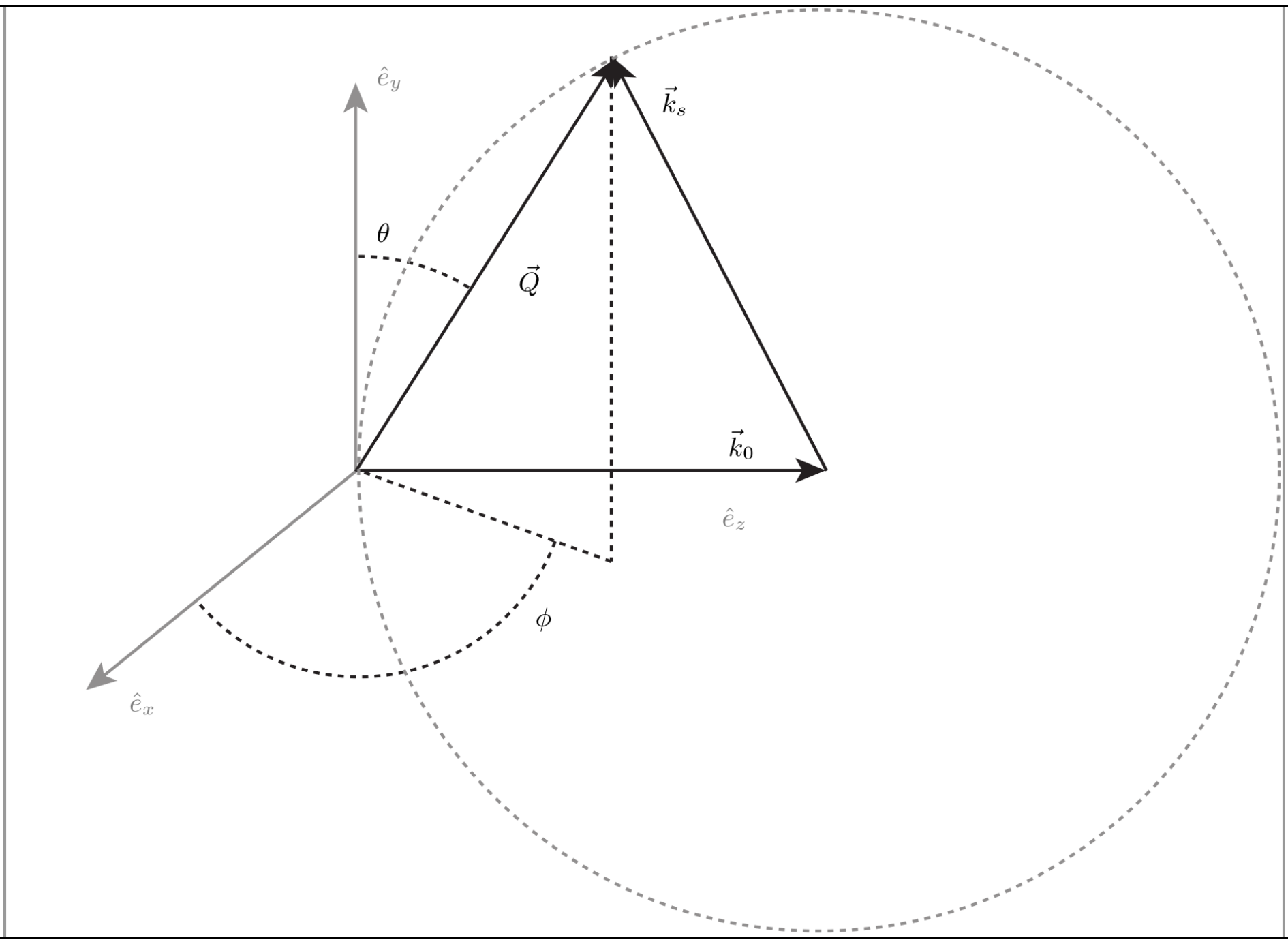

For a linearly polarized pump pulse, the momentum transfer, $\vec{Q}$, is decomposed into its projection onto the $\hat{e}_y$ axis and the $\hat{e}_x - \hat{e}_z$ plane. The large circle depicts the Ewald's sphere, which represents all possible $\vec{Q}$ for an elastic scattering experiment.